\documentclass[prb,twocolumn,altaffilletter,nolongbibliography,numerical,flushbottom,secnumarabic,superscriptaddress,floatfix,nobibnotes]{revtex4-2}

\usepackage{graphicx}
\usepackage{braket}
\usepackage{amsmath, amssymb}
\usepackage{booktabs}  
\usepackage{comment}
\usepackage[normalem]{ulem}
\usepackage{cancel}

\usepackage[breaklinks, pdftex, hyperfootnotes=true, pdfpagelabels, bookmarks, pageanchor]{hyperref}
\hypersetup{%
	colorlinks=true, linktocpage=true, pdfstartpage=1, pdfstartview=FitH, pdfborder={0 0 0},%
	breaklinks=true, pdfpagemode=UseNone, pageanchor=true, pdfpagemode=UseOutlines,%
	plainpages=false, bookmarksnumbered, bookmarksopen=true, bookmarksopenlevel=1,%
	hypertexnames=true, pdfhighlight=/O,
	urlcolor=blue, linkcolor=blue, citecolor=blue,
	}
\usepackage{orcidlink}

\newcommand\identity{1\kern-0.25em\text{l}}

\graphicspath{{figs/}}
 
\usepackage{soul}

\usepackage[colorinlistoftodos]{todonotes}
\presetkeys{todonotes}{inline}{}

\begin{document}
\title{Microwave characterization of superconducting coplanar resonators made out of granular aluminium} 

\newcommand{\affA}{\affiliation{Grupo de Circuitos Cuánticos Bariloche, Div. Dispositivos y Sensores, Centro Atómico Bariloche-CNEA, 8400 San Carlos de Bariloche, Argentina.}}
\newcommand{\affB}{\affiliation{Instituto Balseiro (Universidad Nacional de Cuyo), Bariloche, Argentina.}}
\newcommand{\affD}{\affiliation{Instituto de Nanociencia y Nanotecnolog\'{\i}a (INN), CONICET-CNEA, Argentina.}}

\author{Kelvin J. Ramos}
\affA
\affB
\author{Ivana Curci}
\affA
\affB
\author{Erick Potosí}
\affA\affB
\author{Ignacio Lobato}
\affA\affB
\author{Leonardo Salazar Alarcón}
\affA\affB\affD
\author{Hernán Pastoriza}
\affD
\author{Leandro Tosi}
\email[Corresponding author:]{leandro.tosi@ib.edu.ar}
\affA
\affB
\date{\today}

\begin{abstract}
We have designed, fabricated and characterized microwave resonators made out of thin films of granular aluminium (grAl) with different oxygen content. We extract the contribution of the large kinetic-inductance of this disordered superconductor from the frequency shift of the resonators as a function of temperature, and discuss the correlation with the resistivity of the films. At low temperatures, measurements of the internal quality factor as a function of microwave power indicate the presence of two-level systems, which are inherent to the growth process of the granular aluminium. The characterization methods presented here may be useful for the design of MKIDs, high-impedance resonators and superinductors.
\end{abstract}

\maketitle

\section{Introduction}
High kinetic inductance materials have become increasingly important for a wide variety of applications encompassing superconducting quantum circuits \cite{peltonen2018hybrid,kamenov2020gral,winkel2020implementation,rieger2023gralmonium}, quantum limited amplification \cite{Kher2016_KineticInductanceParametricUpConverter,parker2022degenerate,zhao2023nonlinear,mantegazzini2023high,giachero2023characterization,Splitthoff2024_GateTunableKIPA,zapata2024granular}, high-impedance \cite{janik2025strong,gunzler2025kinetic} and magnetic-field-resilient \cite{Samkharadze2016_HighKineticInductanceNanowireResonators,muller2022magnetic,roy2025highkin} microwave resonators, and also the development of kinetic-inductance photon detectors \cite{day2003broadband,mazin2009microwave,zmuidzinas2012superconducting,day202425}. Since a large kinetic inductance is associated to a reduced density of carriers in the superconducting condensate \cite{schmidt2013physics, tinkham2004introduction}, to a large extent the research is focused on the microwave characterization of the so-called ``disordered superconductors''\cite{Mazin2020_SuperconductingMaterialsMKIDs} including alloys such as TiN \cite{Leduc2010_TiN_Microresonators,Vissers2010_LowLossTiNResonators}, NbN \cite{annunziata2010tunableinductors,Hayashi2014_NbN_Spiral_MKID,carter2019low,niepce2019high,zollitsch2019tuning,mahashabde2020fast,wei2020compact,yu2021magnetic,frasca2023nbn} and NbTiN \cite{xu2019frequency,bretz2022high,yang2024kinetic} and granular materials \cite{moshe2020granular} like granular aluminium \cite{rotzinger2016gral,maleeva2018cqedgral}. Granular aluminium (grAl) is a paradigmatic material that has served as a test-bed to explore the superconducting-insulating transition \cite{deutscher1973gral,levybertrand2019gral} and the role of Josephson coupling in a quasi-2D disordered superconductor in connection to Josephson junction arrays \cite{Pracht2016_EnhancedCooperPairing,maleeva2018cqedgral}. It is obtained by introducing a small partial pressure of oxygen during the deposition of Al via sputtering or evaporation. The resulting films consist of small grains of Al in a matrix of insulating aluminium oxide \cite{deutscher1973granular,ziemann1978oxygen,rotzinger2016gral,levybertrand2019gral,janik2025strong}. The grain size can be controlled with the temperature during growth \cite{deshpande2025tuning} and electrical properties are governed by the percentage of oxygen. The more the Al is oxidized, the larger the sheet resistance and disorder, and in turn, the larger the kinetic inductance. The microwave characterization of superconducting resonators made out of grAl has proved that high quality factors \cite{rotzinger2016gral,zhang2019microresonators,kamenov2020gral,he2021grallumped,gupta2025low}, magnetic field resilience \cite{borisov2020superconducting} and high impedance \cite{gupta2025low,janik2025strong} are within reach. A common issue is the reproducibility in the fabrication of the films, which has been recently addressed with a remote real-time monitoring of the resistance \cite{janik2025strong}. Other concerns are out-of-equilibrium quasiparticles and the role of two-level systems (TLS) limiting the quality factor \cite{grunhaupt2018loss,valenti2018interplay,kristen2023random}. 

In this work we present our measurements of coplanar waveguide (CPW) resonators \cite{pozar2000microwave,simons2004coplanar,goppl2008cpw} made out of grAl films with different oxygen content. We extract the kinetic inductance fraction by fitting the relative frequency shift as a function of temperature with Mathis-Bardeen theory \cite{mattis1958theory}, and compare with the expected value given by the sheet resistance. At low temperatures, we fit the power dependence of the quality factor taking into account the presence of TLS, and obtain relevant parameters that characterize our films. We believe that the characterization methods presented here can be useful to the community of high kinetic inductance devices.

\begin{figure*}[t!]
    \centering
    \includegraphics[width=\linewidth]{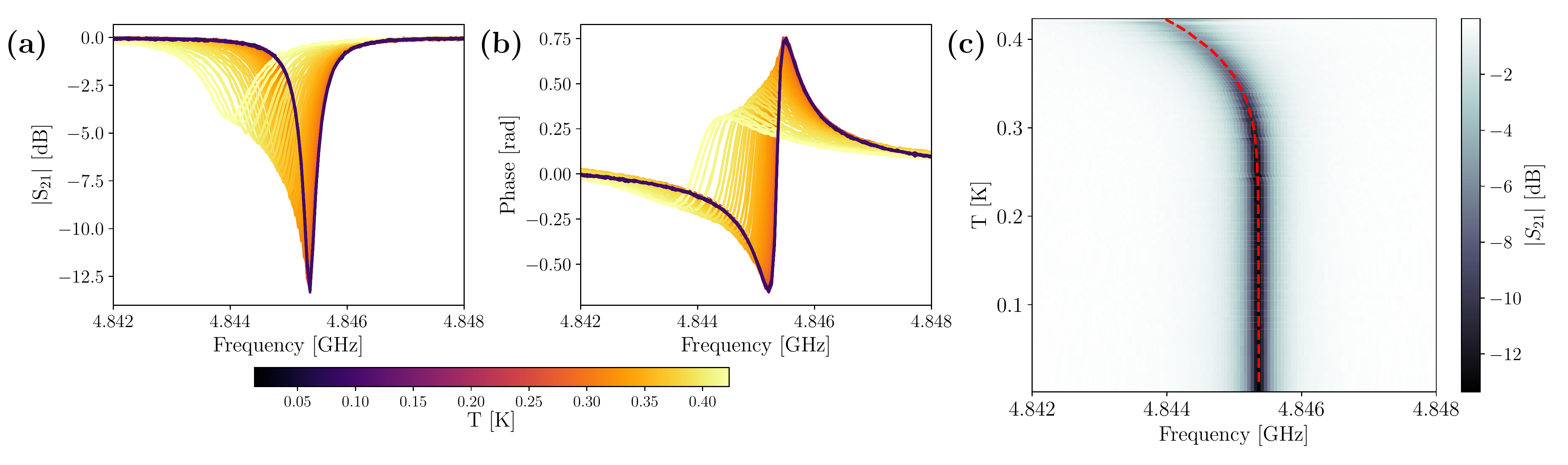}
    \caption{\textbf{Temperature dependence of resonance frequency:}(a) Measured amplitude and (b) phase of the transmission coefficient $S_{21}$ as a function of frequency at various temperatures (indicated by color). (c) $|S_{21}|$ as a function of temperature and frequency in gray-scale map. The dashed line in red represents the fit of resonance frequency using Mattis-Bardeen theory (see text). }
    \label{fig:temperature}
\end{figure*}

\section{Kinetic inductance determination}
Probably the simplest way to extract the kinetic inductance of a thin film is to fabricate a wire with length $l$ and width $w$, measure the sheet resistance at room temperature $R^{\square}_n$ and use the expression \cite{valenti2019gralmkids,annunziata2010tunableinductors}
\begin{equation}
    L^{\square}_{k}=\frac{1}{1.76\pi}\frac{\hbar R^{\square}_n}{k_B T_c},
\label{eq:Lk_res}
\end{equation}
which arises from BCS, valid at temperatures much lower than the critical temperature $T_c$, $T\ll T_c$. The sheet inductance can be increased by decreasing the thickness of the film \cite{lopez2023magnetic}. In this work we follow a different path, which relies on increasing the disorder controlled by the oxidation of the grAl. The estimation provided by this method improves using the sheet resistance at low temperatures (ideally at zero K).  

Another common approach to obtain the kinetic inductance contribution is to fabricate a meander or a coil and directly measure the inductance with an LCR meter \cite{zhdanova2024granular} or embedded in a tank-circuit \cite{weitzel2023BKT}. However, the preferred choice for high-frequency applications is to fabricate microwave resonators on a high-quality substrate, to obtain information about the losses added by the disordered superconductor. For lumped element resonators, the resonance frequency is given by $\frac{1}{2\pi \sqrt{L_t C}}$, where $L_t=L_g+L_k$ is the total inductance (geometric plus kinetic) and $C$ is the geometrical capacitance. The kinetic inductance fraction or ratio can be extracted from the resulting frequency shift
\begin{equation}
    1-\left(\frac{f_0}{f_g}\right)^2=\frac{L_k}{L_g+L_k}\equiv\alpha,
\label{eq:Lk_lumped}
\end{equation}
between the measured frequency $f_0$ and resonance frequency without kinetic inductance $f_g$. This procedure requires prior knowledge of $f_g$ which can be obtained from electromagnetic finite-element simulations or analytic expressions depending on the geometry. In the case of distributed resonators, what matters is the impact of the kinetic inductance on the propagation velocity $v=\frac{1}{\sqrt{\mathcal{L}_t\mathcal{C}}}$ and the characteristic impedance $Z=\sqrt{\frac{\mathcal{L}_t}{\mathcal{C}}}$ determined by the total inductance $\mathcal{L}_t=\mathcal{L}_g+\mathcal{L}_k$ and capacitance $\mathcal{C}$ per unit length \cite{watanabe1994CPWwithki,Porch2005_CoplanarResonators,gao2008thesis}. For a quarter-wavelength resonator of length $\ell$, the resonance frequency is $f_{\text{CPW}}=\frac{v}{4\ell}$, and then,
\begin{equation}
    1-\left(\frac{f_0^{\text{CPW}}}{f_g^{\text{CPW}}}\right)^2=\frac{\mathcal{L}_k}{\mathcal{L}_g+\mathcal{L}_k}=\alpha,
\label{eq:Lk_cpw}
\end{equation}
provides the information on the kinetic inductance, given the measured frequency $f_0^{\text{CPW}}$ and the simulated value $f_g^{\text{CPW}}$. Despite its simplicity, this ``rigid-shift'' method suffers from the problem that the coupling to the read-out circuitry may change with and without kinetic inductance and also displaces the resonance frequency. 

\begin{figure*}[t!]
    \centering
\includegraphics[width=2.1\columnwidth]{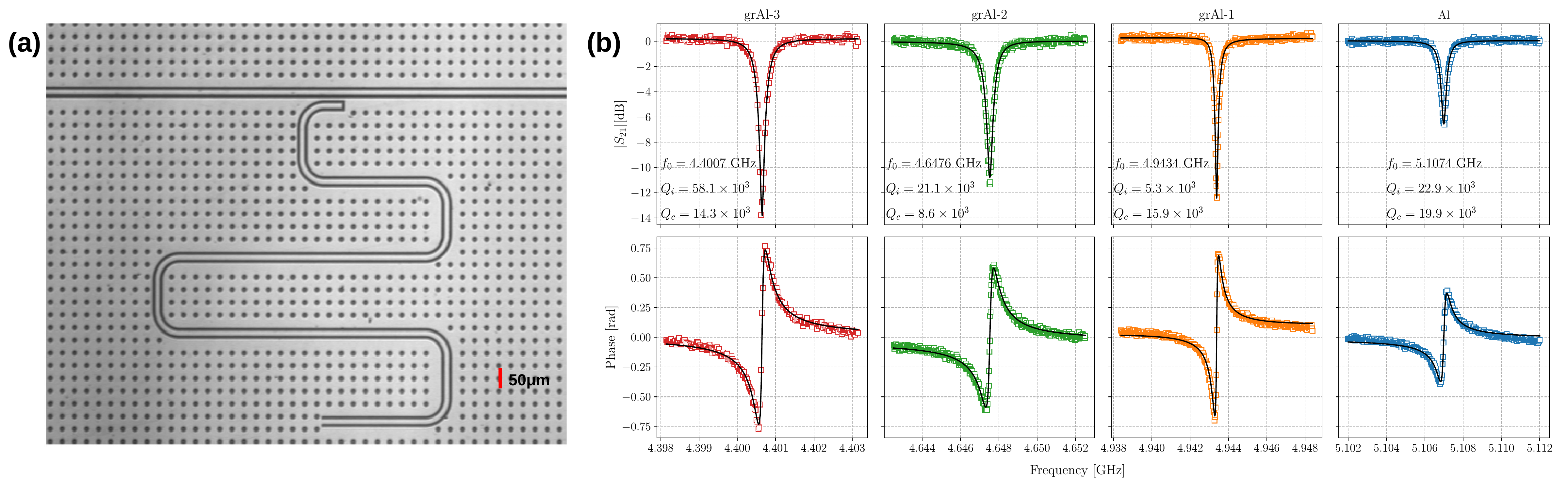}
    \caption{\textbf{GrAl microwave resonators:} (a) CPW $\lambda/4$ resonator coupled to a transmission line fabricated on a grAl film. We present results from devices fabricated on four chips with different oxygen content (blue: Al, orange: grAl-1, green: grAl-2, and red: grAl-3). Three resonators with different length (R1-3) are fabricated per chip. (b) Amplitude (top) and phase (bottom) of the transmission coefficient $S_{21}$ as a function of frequency measured for the resonator R1 in the four samples. The line is the fit using Eq. (\ref{eq:S21}).}
    \label{fig:fig1}
\end{figure*}

A nice alternative is to use the dependence of $L_k$ with temperature \cite{tinkham2004introduction,gao2008thesis}, bias current \cite{annunziata2010tunableinductors} or magnetic field \cite{muller2022magnetic,borisov2020superconducting}. In this work, we focus on the temperature dependence of the resonance frequency, which is probed as we explain below by monitoring the amplitude and phase of the complex transmission coefficient $S_{21}$ (see Fig. \ref{fig:temperature}(a,b)). This dependence can be understood intuitively using a phenomenological two-fluid model \cite{schmidt2013physics}: $L_k(T) \propto 1/n_s(T)$, where $n_s(T)$ is the superconducting density which goes to zero as $T\sim T_c$. So the resonance frequency is expected to decrease with temperature. This can be treated correctly by noting that the surface impedance $Z_s=R_s+iX_s$, for a thin film in the local limit, is related to the complex conductivity $\sigma=\sigma_1-i\sigma_2$ \cite{mattis1958theory,gao2008thesis}, 
\begin{equation}
    Z_s (\omega,T)=\frac{t^{-1}}{\sigma_1(T)-i\sigma_2(T)}
    \label{eq:suface_imped}
\end{equation}
where $t$ is the film thickness. The conductivity can be calculated using Mattis-Bardeen theory \cite{mattis1958theory,Turneaure1991_SurfaceImpedance}.  In the local limit, where the coherence length $\xi_0 \ll \lambda$ (the penetration length), 
\begin{align}
    \frac{\sigma_1(T)}{\sigma_n}&=\frac{4\Delta_0}{\hbar\omega}e^{\frac{-\Delta_0}{k_BT}}\sinh(\xi)K_{0}(\xi),\\
    \frac{\sigma_2(T)}{\sigma_n}&=\frac{\pi\Delta_0}{\hbar\omega}\biggl[1-e^{\frac{-\Delta_0}{k_BT}}\biggl(\sqrt{\frac{2\pi k_BT}{\Delta_0}}-2e^{-\xi}I_{0}(\xi)\biggr)\biggr],
    \label{eq:MB_sigma2}
\end{align}
where $\sigma_n$ is the normal state conductivity, $\Delta_0=1.76 k_B T_c$ is the superconducting gap at $T=0$, $\xi=\hbar\omega/(2k_BT)$, $I_0$ and $K_0$ are the modified Bessel functions of first and second kind, respectively. The change of the resonance frequency $\Delta f=f(T)-f(0)$, and quality factor $\Delta Q_i^{-1}=Q_i^{-1}(T)-Q_i^{-1}(0)$ are then related to the surface impedance through
\begin{align}
    \frac{\Delta f}{f_0}&=-\frac{\alpha}{2}\frac{\Delta X_s}{X_s(0)},\label{eq:MB_fshift}\\
    \Delta Q_i^{-1}&=\alpha\frac{\Delta R_s}{X_s(0)}.
    \label{eq:MB_Qshift}
\end{align}
with $\Delta X_s=X_s(T)-X_s(0)$ and $\Delta R_s=R_s(T)-R_s(0)$.
Thus, by computing $\sigma_{1}(T)$ and $\sigma_2(T)$ and fitting the temperature dependence of the measured resonance frequency and quality factor (see Fig. \ref{fig:temperature}(c)), we can accurately determine the kinetic contribution of the films.

\section{TLS Losses}
TLS originating from impurities, defects and dangling-bonds present at the interfaces (metal-dielectric, metal-air, dielectric-air) are the most well-known cause of microwave losses at low temperatures \cite{muller2019towards,mcRae2020losses,Vallieres2024_LossTangentFluctuations}. They are known to limit the performance of superconducting qubits \cite{Bland2025_2DTransmons} and thus, many strategies have been explored for their mitigation mostly in the direction of reducing surface participation ratio, using more stable oxides, improving the substrate's surface preparation, and the cleaning procedures at every step \cite{OConnell2008_MicrowaveDielectricLoss, Barends2008_ContributionDielectrics,Barends2009_NoiseResonators,Wang2009_CoherenceTimeResonators,Macha2010_LossesCPWResonators,Sage2011_LossSCPWResonators,Wenner2011_SurfaceLossSimulationsCPW,Megrant2012_PlanarSuperconductingResonators,Richardson2016_FabricationArtifacts,burnett2018noise,earnest2018substrate,mcRae2020losses,oh2024lossesNbonSi}. In the case of granular aluminium, where the growth occurs in the presence of oxygen, TLS are unavoidable \cite{kristen2023random} and therefore it is relevant to characterize them to reduce their impact. The coupling to TLS shifts the resonance frequency and changes the quality factor depending on their occupation. This behavior is described by \cite{gao2008thesis,mcRae2020losses} 
\begin{equation}
    \frac{\Delta f}{f_0}=\frac{F\delta_0^{\text{TLS}}}{\pi}\biggl[\text{Re}\biggl\{\Psi\biggr(\frac{1}{2}+\frac{1}{2\pi i}\frac{\hbar \omega_0}{k_BT}\biggr)\biggr\}-\ln \biggl(\frac{\hbar\omega_0}{k_BT}\biggl)\biggr],
    \label{eq:TLS_freq}
\end{equation}
where $\delta^{\text{TLS}}_{0}=1/Q_{\text{TLS}}$ is the TLS loss tangent, $F$ is the filling factor, $\omega_0=2\pi f_0$ and $\Psi$ is the di-Gamma function; and
\begin{equation}
    \frac{1}{Q_i}=F\delta^{TLS}_{0}\frac{\tanh\biggl(\frac{\hbar \omega_0}{2k_BT}\biggr)}{\sqrt{1+\left(\frac{\langle n \rangle}{n_c}\right)^{\beta}}}+\delta_1,
    \label{eq:TLS_Q}
\end{equation}
for the internal quality factor $Q_i$, where $\langle n \rangle$ is the average photon number in the resonator,
\begin{equation}
    \langle n \rangle = 4P_{in}\frac{Q^2}{Q_{c}\hbar\omega_0^2},
    \label{eq:n_phot}
\end{equation}
for a given microwave power $P_{in}$ at the input of the resonator and 
$Q$ is the total quality factor,
\begin{equation}
\frac{1}{Q}=\frac{1}{Q_i}+\frac{1}{Q_{c}},
\label{eq:total_Q}
\end{equation}
with $Q_c$ the coupling quality factor. $n_c$ is a characteristic number of photons that generate the electric field required to saturate the TLS, $\beta$ is an empirical parameter describing the interaction with the TLS, and $\delta_1$ includes all the remaining losses. 

\section{Resonator fabrication}
\subsection{Granular aluminium film growth}
The granular aluminium films characterized in this work were grown by sputtering high-purity aluminium onto high-resistivity silicon substrates at room temperature. The Si wafer was cleaned with a buffered oxide etch (BOE) prior to deposition in order to remove any native Si oxide layer. The pressure in the chamber before the entry of argon and oxygen flow was in the range of $(9.0\pm0.1)\times 10^{-7}$ mbar. The oxygen flow, together with the deposition rate, determines the resistivity of the film. The details on the film growth and characterization can be found in App. \ref{ap:films_growth}. 

We present results from devices fabricated on four different chips whose properties are summarized in Table \ref{tab:films}. From the $T_c$ (see measurement details in App. \ref{ap:extra}) and the sheet resistance we obtain the expected value of the sheet kinetic inductance from Eq. (\ref{eq:Lk_res}). 

\begin{table}[!ht]
    \centering
    \begin{tabular}{|c|c|c|c|c|c|}
    \hline \hline
        Sample & $t$ [nm] & $T_c$ [K] & $R^{\square}$ [$\Omega/\square$] & $\rho$ [$\mu \Omega$.cm] & $L^{\square}_k$ [pH/$\square$]  \\ \hline\hline
        Al & 63.0 & 1.44 & 2.48 & 15.62 & 2.37 \\ 
        grAl-1 & 73.1& 1.64 & 8.87 & 64.84 & 7.46\\ 
        grAl-2 & 65.5 & 1.76 & 13.75 & 90.06 & 10.79\\ 
        grAl-3 & 57.5 & 1.73 & 4.19 & 24.10 & 3.35                                 \\ \hline \hline
    \end{tabular}
    \caption{Relevant parameters of the grAl films characterized in this work. See also App. \ref{ap:extra}. The value of last column is obtained from Eq. (\ref{eq:Lk_res}).}
    \label{tab:films}
\end{table}

\subsection{Microwave resonators design and microfabrication}
In each grAl film indicated in Table \ref{tab:films} we fabricate three $\lambda/4$ resonators \cite{pozar2000microwave, goppl2008cpw} (named R1, R2 and R3) of different lengths, coupled to the same transmission line in a hanger configuration. The coupling of each one is adjusted to get the same total quality factor. Figure \ref{fig:fig1}(a) shows one of our CPW resonators fabricated using standard optical lithography and wet chemical etching. The central conductor width $w=20\ \mu$m and the gap to the ground plane $g=11\ \mu$m, are chosen to give a 50$\Omega$ impedance in absence of kinetic inductance. Further details on the design of the CPW resonators can be found in App. \ref{ap:res_design}.

\subsection{Measurement setup and parameter extraction}
The chips are bonded to a sample holder and subsequently mounted at the mixing chamber stage of a dilution refrigerator. Measurements of the complex transmission coefficient $S_{21}$ are carried out in the range from base temperature ($\sim$10 mK) to above $T_c$ using a vector network analyzer (VNA). The input signal is heavily attenuated ($-70$dB) and the output signal is amplified by using a cryogenic HEMT amplifier at the 4K stage, followed by additional amplification at room temperature (see App. \ref{ap:meas_setup}). The resonance frequency and quality factor are extracted by fitting the data with \cite{probst2015efficient} 
\begin{equation}
    S_{21}(f)=ae^{i\phi_0}e^{-2\pi if\tau}\biggl( 1-\frac{Q/Q_ce^{i\phi}}{1+2iQ(f/f_0-1)}\biggr)
    \label{eq:S21}
\end{equation}
where $a$ is the background transmission, $f_0$ is the resonant frequency, $Q$ and $Q_c$ are the total and coupling quality factors, respectively, whereas $\phi_0$ is an offset phase, $\tau$ is the line delay and $\phi$ takes into account any additional impedance mismatch. Figure \ref{fig:fig1}(b) shows the amplitude and phase of $S_{21}$ for resonator R1 measured at the base temperature with  $\langle n\rangle\sim \times 10^3$ for the four different samples (i.e. same geometry in every case): Al, grAl-1, grAl-2 and grAl-3. The markers correspond to the experimental points while the line is the fit with Eq. (\ref{eq:S21}). 

\section{Characterization of grAl resonators}  
\begin{figure}[t!]
    \centering
\includegraphics[width=\columnwidth]{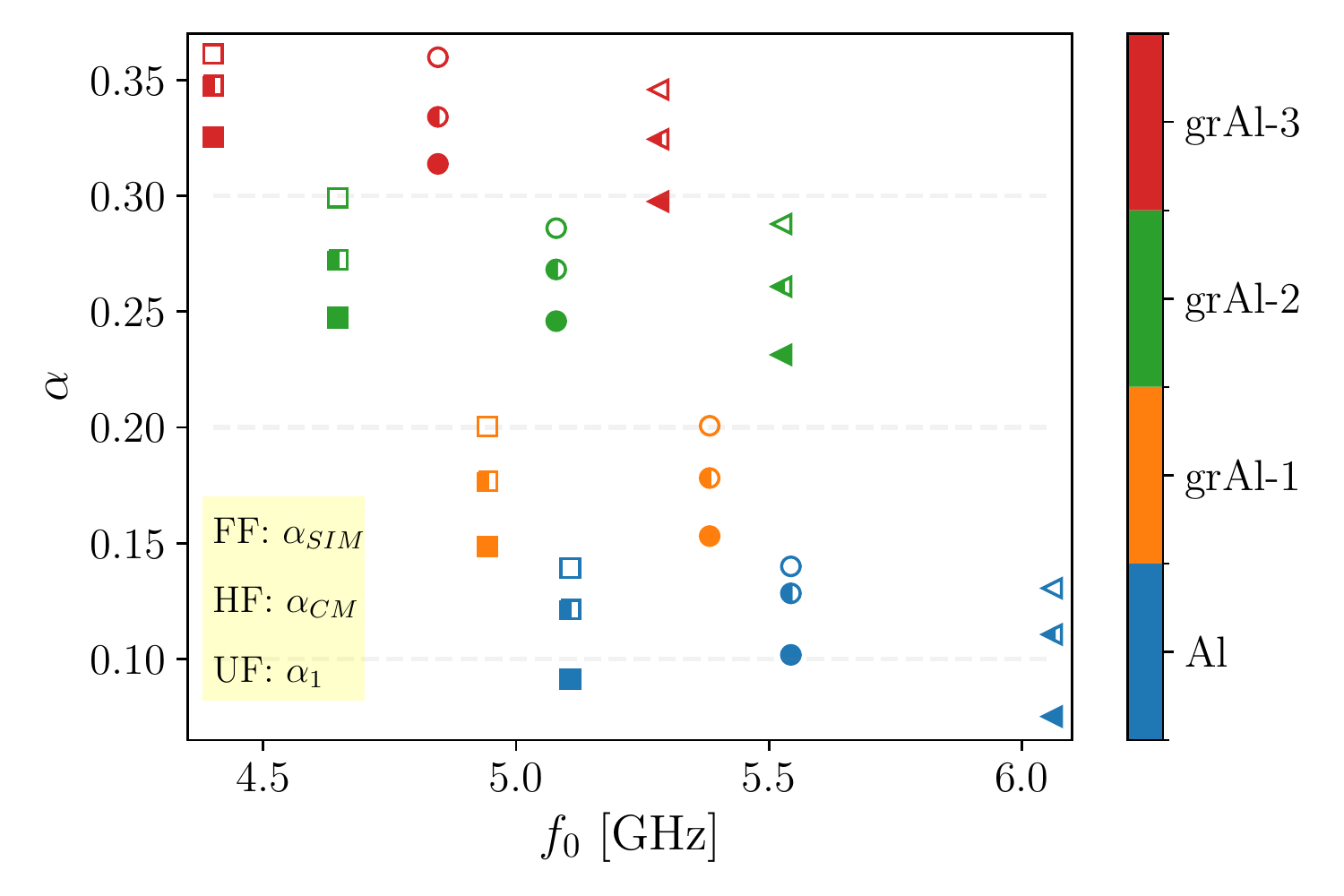}
    \caption{\textbf{Comparison between different methods:} Kinetic inductance fraction as function of measured resonance frequencies for resonators R1-3 in each sample. Values are obtained from rigid shift using Eq. (\ref{eq:Lk_cpw}) where the geometric resonant frequency is calculated from either from finite-element simulations (yielding $\alpha_{\text{SIM}}$, fully-filled (FF) markers) or from the conformal mapping technique (yielding $\alpha_{\text{CM}}$, Half-Filled (HF) markers). $\alpha_1$ is obtained from the fit of the frequency shift with Mattis-Bardeen (Un-Filled (UF) markers). Colors represent different samples - blue: Al, orange: grAl-1, green: grAl-2, and red: grAl-3 - while symbols denote resonators: square (R1), circle (R2) and triangle (R3).}
    \label{fig:fig2}
\end{figure}
\begin{figure}[t!]
    \centering
\includegraphics[width=\columnwidth]{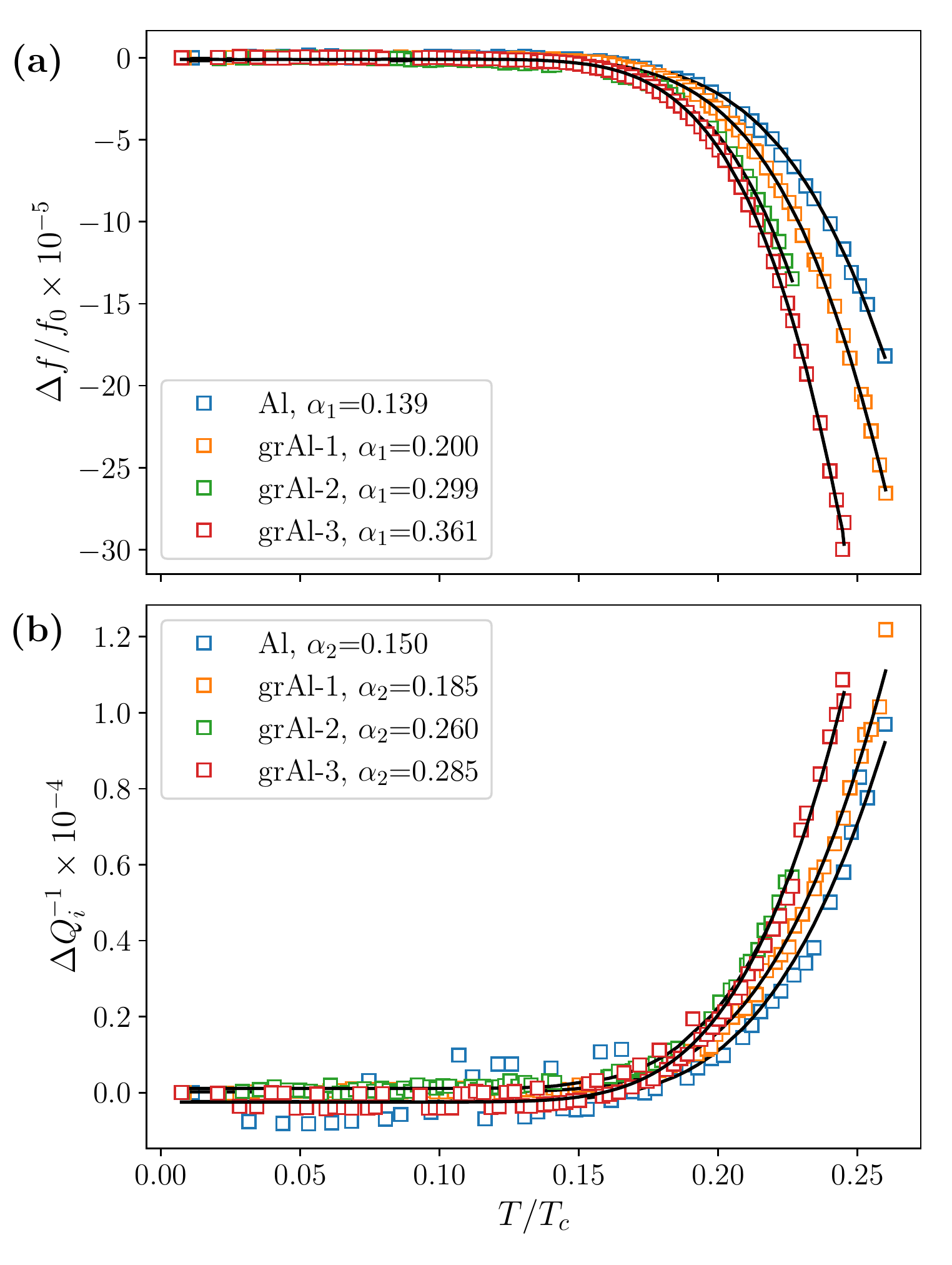}
    \caption{\textbf{Mattis-Bardeen fit of temperature dependence:} (a) Relative frequency shift and (b) Inverse quality factor shift as functions of $T/T_c$. Solid lines correspond to fits with Eqs. (\ref{eq:MB_fshift}) and (\ref{eq:MB_Qshift}), based on Mattis-Bardeen theory. The extracted values of $\alpha$ in each case are indicated in the legend.}
    \label{fig:films}
\end{figure}

\subsection{Kinetic inductance fraction}
From the fit in Figure \ref{fig:fig1}(b), taking the Al film as a reference ($f_0=5.1074$ GHz), a systematic downward shift in resonance frequency is observed: -164\ MHz for grAl-1, -459.8\ MHz for grAl-2 and -706.7\ MHz for grAl-3. As discussed above, the kinetic ratio $\alpha$ can be extracted from the observed frequency shifts, i.e the rigid shift, in conjunction with Eq. (\ref{eq:Lk_cpw}). The value of the resonance frequency in absence of kinetic inductance can be obtained by two methods (see App. \ref{ap:res_design}): by finite-element simulations, yielding $\alpha_{\text{SIM}}$, and by analytic result from Conformal Mapping Technique (CMT), which yields $\alpha_{\text{CM}}$. In this case we calculate the frequency as $f_{g}^{\text{CM}}=(4\ell\sqrt{\mathcal{C}_g \mathcal{L}_g})^{-1}$ using Eq. (\ref{eq:Cg}) and (\ref{eq:Lg}). For resonators with lengths: $\ell_{R_1}=5.42697$\ mm, $\ell_{R_2}=4.97984$\ mm and $\ell_{R_3}=4.60149$\ mm, the simulated resonance frequencies $f_{g}^{\text{SIM}}$ are: 5.35717\ GHz, 5.84901\ GHz, 6.301458\ GHz, respectively. The frequencies obtained by using CMT are: 5.44832\ GHz, 5.93752\ GHz, 6.42571\ GHz. The CMT-derived frequencies closely match the simulated frequencies, which in addition take into account the correction due to the coupling capacitor to the transmission line. The resulting values of $\alpha_{\text{SIM}}$ and $\alpha_{\text{CM}}$ are plotted in Fig. \ref{fig:fig2} as functions of the measured resonance frequencies, using fully-filled (FF) and half-filled colored markers (HF), respectively. The average values for the kinetic inductance fraction extracted using the simulated frequency are $\bar{\alpha}_{\text{SIM}}(\text{Al})=(0.09\pm	0.01)$, $\bar{\alpha}_{\text{SIM}}(\text{grAl-1})=(0.151\pm	0.003)$, $\bar{\alpha}_{\text{SIM}}(\text{grAl-2})=(0.242\pm	0.009)$ and $\bar{\alpha}_{\text{SIM}}(\text{grAl-3})=(0.31\pm	0.01)$. When using the frequency expected from the conformal mapping equations we obtain $\bar{\alpha}_{\text{CM}}(\text{Al})=(0.120\pm	0.009)$, $\bar{\alpha}_{\text{CM}}(\text{grAl-1})=(0.177\pm	0.001)$, $\bar{\alpha}_{\text{CM}}(\text{grAl-2})=(0.267\pm	0.006)$ and $\bar{\alpha}_{\text{CM}}(\text{grAl-3})=(0.335\pm	0.012)$. The kinetic inductance fraction is in the order of 30\% for the grAl-3 sample.

These previous results can be contrasted with the fit of the frequency and quality factor shifts using Mattis-Bardeen theory, with Eqs. (\ref{eq:MB_fshift}) and (\ref{eq:MB_Qshift}), which yield $\alpha_1$ and $\alpha_2$, respectively. The relative shifts for R1 are shown in Fig. \ref{fig:films}, the fits are plotted with lines. In the fit, the $T_c$ is fixed to the measured value for each film (see App. \ref{ap:extra}). We have found that in general $\alpha_1$ and $\alpha_2$ do not coincide, with a difference ranging from 1 to 22\%. We believe that this might be due to an additional contribution to the shift of the quality factor with temperature. In order to compare to the other methods, we adopt $\alpha_1$ as our estimate of the kinetic inductance fraction, as the relative frequency shift provides a more reliable determination of $\alpha$. The values (see the unfilled (UF) colored markers in Fig. \ref{fig:fig2}) are of the same order as those extracted from the rigid shift, although systematically higher. The dispersion among the values for R1-3 in the same chip is also larger than the error of the fit, we obtain $\bar{\alpha}_1(\text{Al})=(0.137\pm	0.005)$, $\bar{\alpha}_1(\text{grAl-1})=(0.201\pm	0.001)$, $\bar{\alpha}_1(\text{grAl-2})=(0.291\pm	0.007)$ and $\bar{\alpha}_1(\text{grAl-3})=(0.356\pm	0.008)$. We observe that the results extracted from Mattis-Bardeen are on average 25\% larger than the kinetic inductance fraction extracted from the rigid shift using the simulated frequency and 9\% using the analytic conformal mapping expressions. 

We can compute the total kinetic inductance of the resonators through $L_k=\alpha L_g/(1-\alpha)$, where $L_g=1/(C_g^{}\omega_g^2)$, $C_g=\pi/(4\omega_g Z_0)$ and $\omega_g=2\pi f_g^{\text{SIM}}$\cite{pozar2000microwave}. Results are summarized in Table \ref{tab:films_v2}. If we consider only the central conductor of R1, $\ell/w= 5427/20\approx 270\ \square$, then a naive estimate of the sheet inductance would be $L_{k}^{\square}\sim 1.13, 1.75, 2.99$ and $3.96$\ pH/$\square$, for the different samples. But this in general estimates wrongly the value since in the CPW geometry a different number of squares may contribute, depending on the aspect ratio and the thickness. One common strategy is to solve Maxwell and London equation simultaneously using a finite-element solver, in particular when $t\le \lambda_{\text{eff}}$ and no analytic expressions are available \cite{gao2008thesis}, being $\lambda_{\text{eff}}=\frac{\lambda^2}{t}$ the effective penetration depth, and $\lambda$ the bulk penetration depth. We followed a different procedure: using conformal mapping expressions which include the kinetic inductance \cite{watanabe1994CPWwithki} (see Eq. (\ref{EqG}) from App. \ref{ap:res_design}) we calculated the $\alpha(L_{k}^{\square})$ and solved for $\alpha (L_{k}^{\square})=\alpha_{1}$, as obtained in the experiment. The average values are $L_k^{\square}(\text{Al})=(1.8\pm0.5)$ pH/$\square$, $L_k^{\square}(\text{grAl-1})=(2.33\pm0.02)$ pH/$\square$, $L_k^{\square}(\text{grAl-2})=(5.4\pm0.2)$ pH/$\square$ and $L_k^{\square}(\text{grAl-3})=(6.9\pm0.3)$ pH/$\square$. For future applications, they could be strongly increased by growing thinner films. We observe that the number of squares ($\#_\square$, as reported in the last column of Table \ref{tab:films_v2}) that contribute to kinetic inductance is $\sim 0.5 (\ell/w)$ of the central conductor. Moreover, the sheet inductance obtained this way is of the same order of magnitude but differs significantly from the prediction in Eq. (\ref{eq:Lk_res}). It is unclear to us why the difference is so large, but the observations regarding the kinetic inductance fraction are more consistent with the value extracted from Mattis-Bardeen. One important point is that $L_k^\square=\mu_0 \lambda^2/t\equiv \mu_0 \lambda_{\text{eff}} \Rightarrow \lambda_{\text{eff}}\approx 800\ $nm, for 1\ pH/$\square$ so the use of the theoretical expressions to extract $\alpha$ is not justified and we take these values only as indicative.    

\begin{table}[t!]
\begin{tabular}{|c|c|c|c|c|c|}
\hline \hline 
sample& $\alpha_1$ & $\alpha_2$ & $L_{k}$[nH] & $L_k^{\square}$[pH$/\square$] & $\#_\square$ \\ \hline \hline
Al, R1 & 0.139 & 0.150  & 0.306  & 2.104 & 145\\
Al, R2 & 0.140 & 0.133  & 0.282  & 2.121 & 133\\
Al, R3 & 0.131 &  0.140  & 0.241  & 1.194 & 202\\ \hline

grAl-1, R1 & 0.200 &  0.185  & 0.474  & 3.480 & 136\\
grAl-1, R2 & 0.201 &  0.22   & 0.435  & 3.502 & 124\\ 
grAl-1, R3 &  -    &  -      &  -     &   -     &-    \\ \hline
       
grAl-2, R1 & 0.299 & 0.26   & 0.807  & 5.653 & 142\\
grAl-2, R2 & 0.286 &  0.243  & 0.694  & 5.308 & 130\\
grAl-2, R3 & 0.288 & 0.224  & 0.650  & 5.361 & 121\\ \hline

grAl-3, R1 & 0.361 & 0.285  & 1.070  & 7.083 & 151\\
grAl-3, R2 & 0.360  & 0.353  & 0.974  & 7.053& 138\\
grAl-3, R3 & 0.346  & 0.341  & 0.850  & 6.633& 128\\ \hline \hline       
\end{tabular}
\caption{Kinetic inductance contribution extracted from Mattis-Bardeen fit: $\alpha_1$ obtained from fitting the frequency shift with Eq. (\ref{eq:MB_fshift}) and $\alpha_2$ the quality factor shift with Eq. (\ref{eq:MB_Qshift}). The $L_k$ is obtained from $\alpha_1$ using the simulated geometrical inductance. The sheet kinetic inductance is obtained from matching the theoretical $\alpha(L_k^{\square})$ to the experimental value $\alpha_1$. The number of squares is $L_k/L_k^{\square}$ and provides an idea of the geometrical correction.}
\label{tab:films_v2}
\end{table}

\begin{figure*}[t!]
    \centering
\includegraphics[width=2\columnwidth]{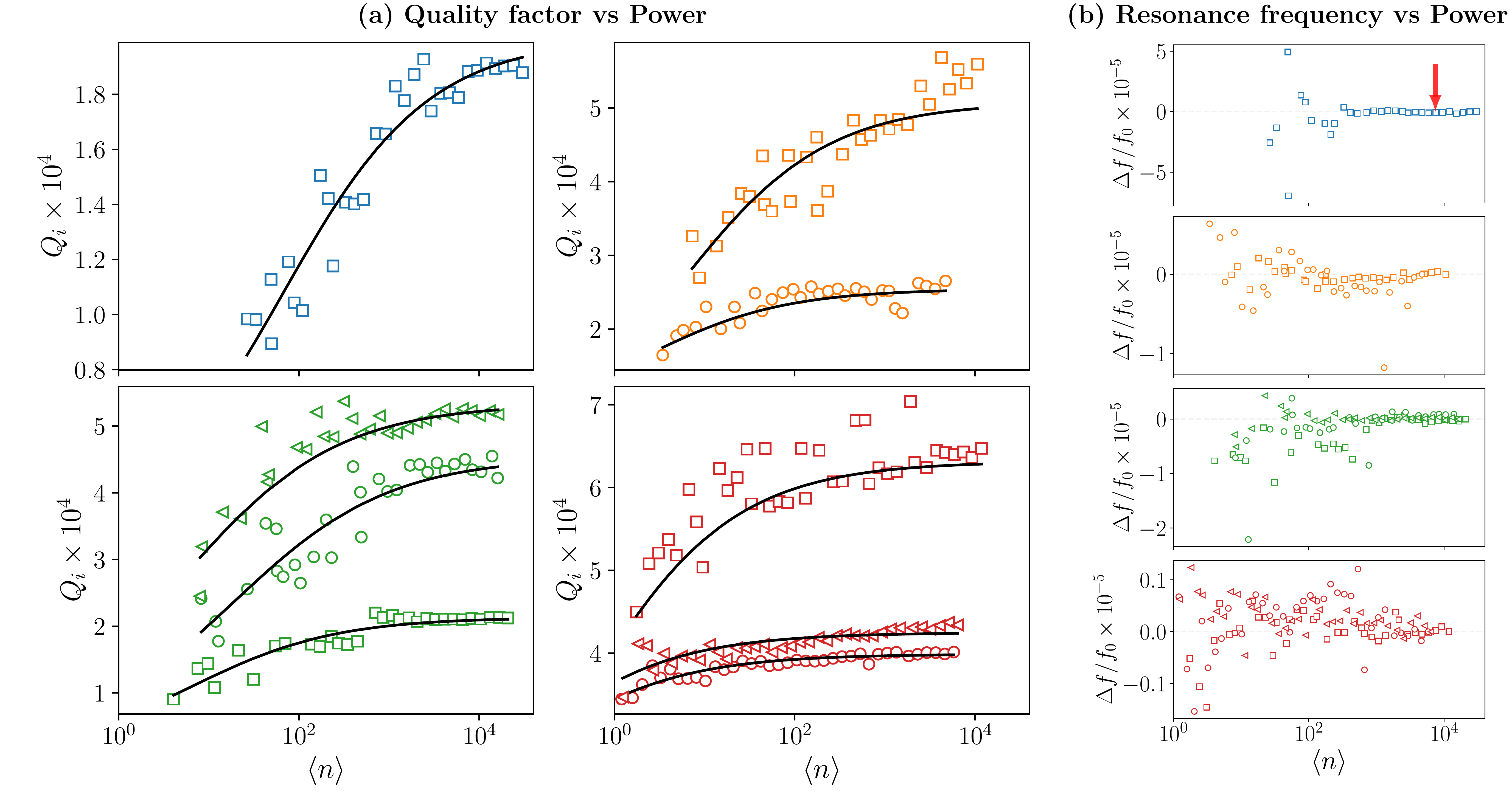}
\caption{\textbf{TLS losses:} (a) Internal quality factor as a function of the number of photons measured at $T=10\ $mK for the different resonators in each sample (blue: Al, orange: grAl-1, green: grAl-2, and red: grAl-3). The lines correspond to fits with Eq. (\ref{eq:TLS_Q}). (b) Power dependence of the resonance frequency before the onset of non-linearity. Here, $f_0$ is taken in a regime of average photon number $\langle n \rangle \sim  10^3$, where TLS are expected to be saturated.}
\label{fig:fig3}
\end{figure*}

\subsection{TLS losses}
\begin{figure}[h!]
    \centering
    \includegraphics[width=0.8\linewidth]{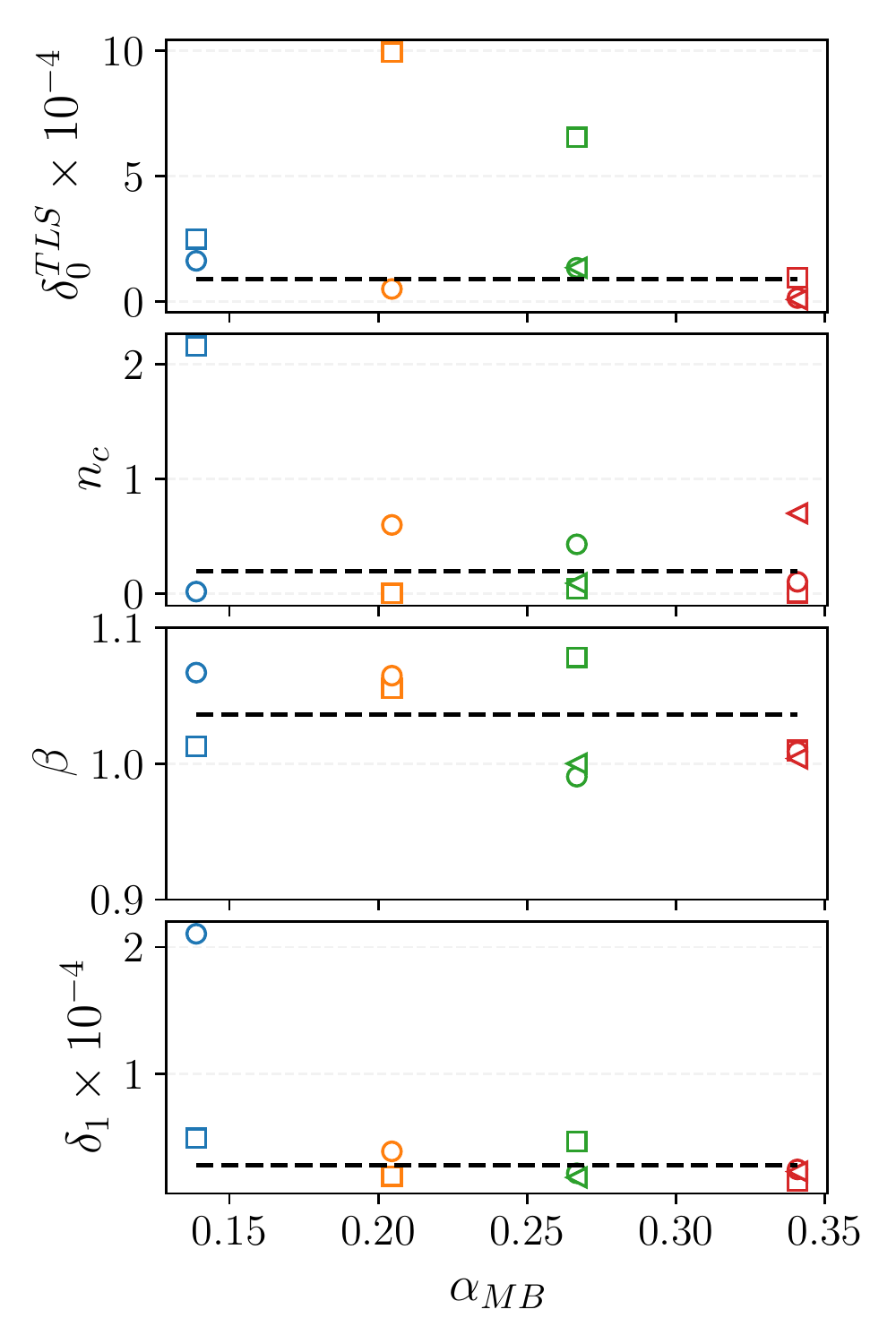}
    \caption{\textbf{TLS contribution:} From top to bottom: $F\delta_{\text{TLS}}$, $n_c$, $\beta$ and $\delta_{1}$ extracted from the fit of data in Fig. \ref{fig:fig3} using Eq.  (\ref{eq:TLS_Q}) (Colors represent different samples - blue: Al, orange: grAl-1, green: grAl-2, and red: grAl-3 - while symbols denote resonators: square(R1), circle(R2) and triangle(R3)). }
    \label{fig:fig5}
\end{figure}

We focus now on the analysis of microwave losses in our grAl resonators. In Fig. \ref{fig:fig3}(a) we show the internal quality factor and in (b) the resonance frequency shift both as a function of the average photon number, measured at base temperature for the different resonators in each sample. The increase of $Q_i$ is associated with the presence of two-level systems that couple to the electromagnetic mode of the resonators and get saturated as the microwave power is increased \cite{gao2008thesis,mcRae2020losses}. We fit the measurements with Eq. (\ref{eq:TLS_Q}) (see lines in (a)) and extract the parameters $\beta$, $n_c$, $F\delta_{\text{TLS}}$ and $\delta_{1}$. These are shown in Fig. \ref{fig:fig5}. The contribution of the TLS determined by $F\delta_{\text{TLS}}$ is in average 1.5$\times 10^{-4}$, with two outliers resonators which show a larger participation ratio. This value is similar to what has been reported in quarter wavelength CPW resonators \cite{mcRae2020losses}. We do not observe a clear systematic trend with the kinetic inductance fraction. The values of $\beta$ are very close to 1, which would be expected from the microscopical theory of the resonator-TLS coupling \cite{Wang2009_CoherenceTimeResonators,gao2008thesis}. The saturation power is below one-photon. The non-TLS losses would limit the quality factor to around $5\times 10^{4}$. Comparing Fig. \ref{fig:fig3}(a) and (b), we observe that at the value of $\langle n \rangle$ (indicated by a downward red arrow) used for the temperature dependence in Fig. \ref{fig:films}, the effect of TLS on the frequency shift is much weaker than on the internal quality factor. We think that since the TLS are not fully saturated this contributes to the difference between the extracted values of $\alpha_1$ and $\alpha_2$ with Mattis-Bardeen.  
\section{Conclusion and Discussion}
We have designed and fabricated CPW microwave resonators made out of granular aluminium thin films with different oxygen content. From the microwave characterization we were able to extract the kinetic inductance contribution which combined with finite-element simulations provides the value of the total kinetic inductance. Using conformal mapping expressions we were able to extract the sheet kinetic inductance to compare with the simple expectation from the sheet resistance. Additionally, we characterized the contribution of two-level systems to the microwave losses at low temperatures and did not find a clear correlation with the level of oxidation of the Al films. We believe the methods presented here can be useful to the community exploring high kinetic inductance materials for high-frequency applications.

\appendix
\counterwithin{figure}{section}

\section{Granular Aluminium films material characterization}
\label{ap:films_growth}

\begin{figure*}[t!]
    \centering
\includegraphics[width=\linewidth]{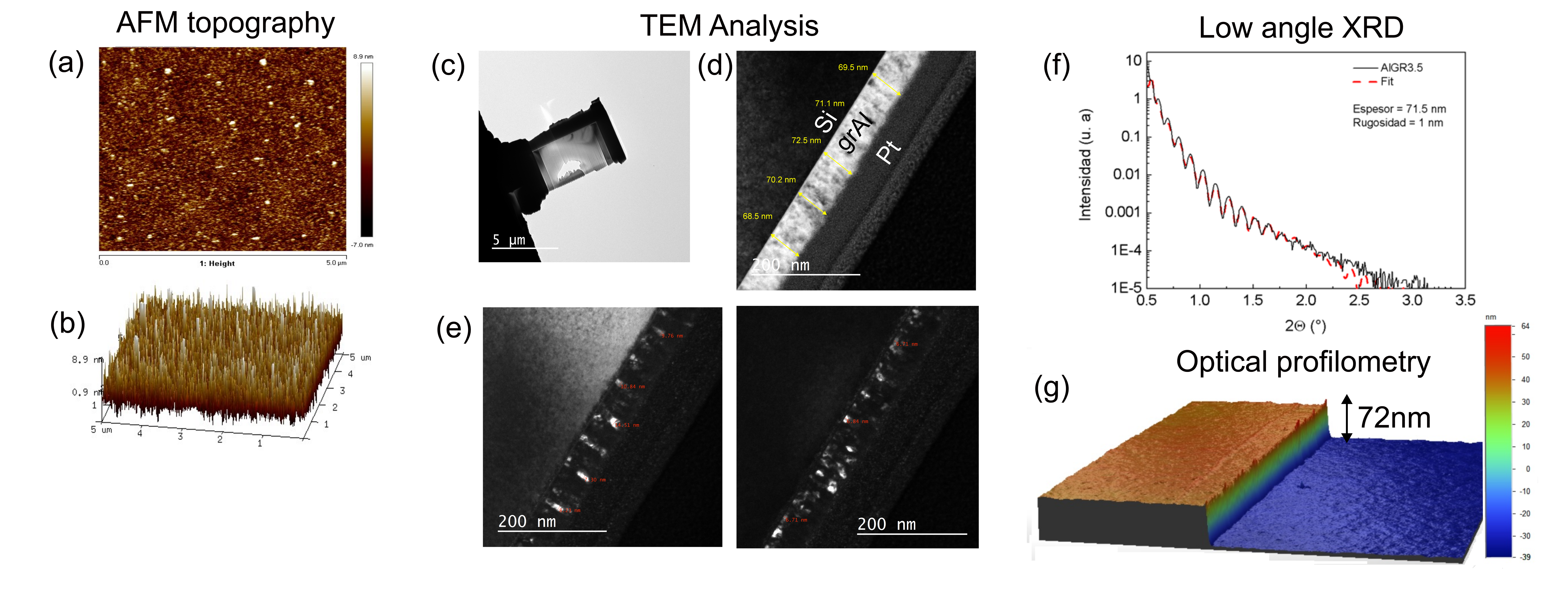}
    \caption{\textbf{Characterization of sputtered grAl films:} (a) AFM topography 10x10\ $\mu$m map and (b) 3D view of a grAl film with nominal 3.2\% of oxygen. (c) GrAl slab produced by FIB. (d) Bright-field TEM image of a 3.1\% grAl film (the indicated regions correspond to: grAl, Si and the protecting Pt layer). (e) First ring of the diffraction pattern at the grAl in a dark-field mode at two different regions of the film. (f) Low-angle refraction pattern (intensity vs 2$\theta$) for a grAl film with 3.5\% of oxygen. The fit is performed using the software \textit{Parratt 1.6}. (g) 3D topographic image of the profile obtained after the chemical etching of a part of a film with 3.2\% of oxygen. The step height obtained from the average profile is (72 $\pm$ 2) nm. }
    \label{fig:fig4}
\end{figure*}

The growth of granular Al thin films was performed at the Clean room facility at Centro Atómico Bariloche in a Alcatel sputtering machine. We used a 99.95\% purity Al circular target of 3.8 cm diameter and 10 x 10\ mm or 15 x 15\ mm Si substrates. The reference pressure was $\sim 10^{-7}$ mbar, prior to the flow of Ar which is controlled by a needle valve. By growing Al films at different pressures and applied voltages, we determined the working pressure (10 mTorr) and the power (100 W) that minimize the roughness of the films \cite{curci2022tesis}. The oxygen partial pressure is determined with respect to Ar, in the following we refer to the films quoting the nominal percentage. This must be taken as indicative though, since we have observed a large dispersion in the properties for the same nominal oxygen percentage. This is a well-known problem of grAl \cite{janik2025strong} and this is why in the main text we refer only to the sheet resistance.

The roughness of the films was characterized with AFM measurements. A topography 10x10\ $\mu$m map of a grAl film with 3.2\% of oxygen is shown in Fig. \ref{fig:fig4} (a) with a 3D view in (b). The extracted root mean roughness $R_{ms}$ for films with different oxygen content in average is (1.2 $\pm$ 0.3)\ nm and does not change significantly with the oxygen percentage. In a pure Al film we obtained $\bar{R}_{ms}$=(1.1 $\pm$ 0.4)\ nm, indicating that the growth conditions are well optimized.

Another relevant morphology property of the films is the grain distribution. We took many SEM images under different conditions but were not able to distinguish the Al grain in the AlOx matrix. An alternative approach is to use TEM: we used a FIB to produce thin slabs (see Fig. \ref{fig:fig4}(c)) which we can then image with TEM. A bright-field image of a 3.1\% grAl film is presented in panel (d), where the grAl, Si and the protecting Pt layers are indicated. The information on the grain size can be obtained from the diffraction pattern at the grAl in a dark-field mode, as shown in (e) for the first ring of the diffraction pattern at two different regions of the films. The size distribution ranges from 3\ nm up to tens of nm, with grains in a columnar morphology oriented in the growth direction. From the distribution histograms obtained using \textit{ImageJ}, we obtain an average size $(6\pm 1)$\ nm, which is in agreement with the values reported in the literature for growth at room temperature \cite{levybertrand2019gral,deshpande2025tuning,janik2025strong}. 

The bright field TEM image (Fig. \ref{fig:fig4}(d)) provides also information on the thickness of the film, resulting (70 $\pm$ 2) nm. In combination with the growth time of 3 minutes, we get our growth rate to be $\sim$24 nm/min. As shown in Fig. \ref{fig:fig4}(f), we have also determined the thickness by using low-angle XRD and (g) optical profilometry. In (f) a typical example of the low-angle refraction pattern (intensity vs 2$\theta$) for a grAl film with 3.5\% of oxygen is shown, together with the fit performed using the software \textit{Parratt 1.6}. From fitting we obtain the thickness $(71.5\pm 0.5)$\ nm and an average roughness of 1\ nm. In (g) we show a 3D topographic image of the profile obtained after the chemical etching of a part of a film with 3.2\% of oxygen. The etching solution consists of 16 parts $\text{H}_3\text{PO}_4$, 2 parts $\text{HNO}_3$, 1 part $\text{CH}_3\text{COOH}$ and 1 part $\text{H}_2\text{O}$. The step height obtained from the average profile is (72 $\pm$ 2)\ nm, so that the different methods give compatible results.

\section{Design of microwave resonators}
\label{ap:res_design}

\begin{figure*}[t!]
    \centering
    \includegraphics[width=\linewidth]{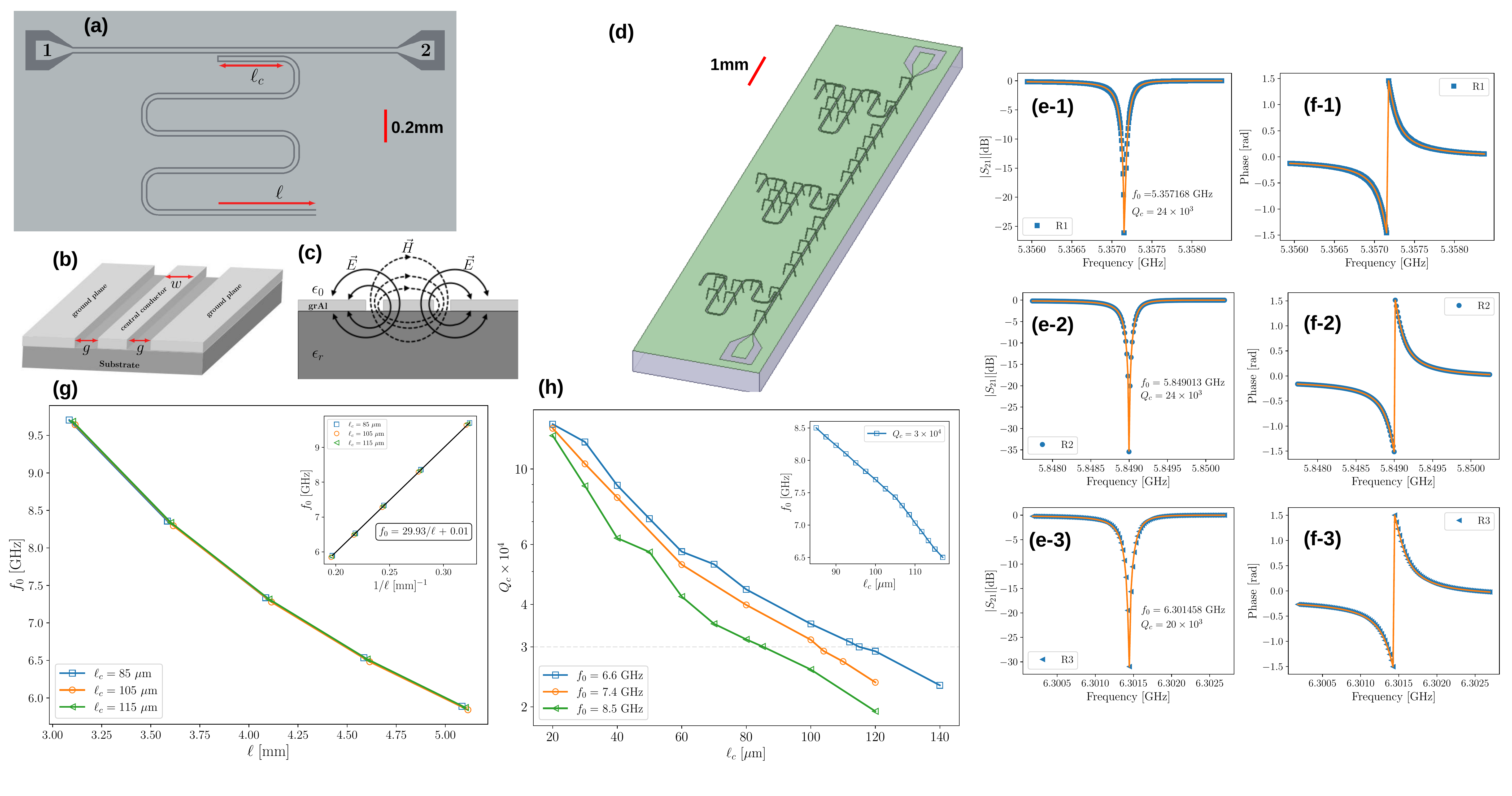}
    \caption{\textbf{Design of CPW microwave resonators:}(a) Quarter-wavelength resonator of length $\ell$ coupled to a transmission line over a coupling length $\ell_c$. (b) Top view of the CPW structure, showing the central conductor width $w$ and gap $g$ on a dielectric substrate. (c) Cross-sectional view of the CPW transmission line illustrating the propagation of the quasi-TEM electromagnetic fields. (d) Design of three $\lambda/4$ microwave resonators with different $\ell$ and $\ell_c$ (tuned to achieve the same coupling quality factor), created by using qiskit metal \cite{qmetal} and prepared for simulation in Ansys HFSS \cite{ansys}. Simulated (e-1, e-2, e-3) amplitude and (f-1, f-2, f-3) phase for the three designed $\lambda/4-$ resonators. The lines are fits with Eq. (\ref{eq:S21}). (g) Extracted resonance frequency for different $\ell$; the inset show a linear fit of frequency versus inverse length. (h) Coupling quality factor for different values of $\ell_c$ for three fixed frequencies ($\ell$ is modified to keep the frequency fixed). The inset shows additional points for the coupling length which yields a target $Q_c$=30000 at different resonance frequencies.}
    \label{fig:fig6}
\end{figure*}

We work with distributed quarter-wave-length microwave resonators made out of coplanar waveguides (CPW)\cite{pozar2000microwave,simons2004coplanar,Porch2005_CoplanarResonators,goppl2008cpw,ramosvillalobos2022tesis} as shown in Fig. \ref{fig:fig6}(a-c). For a CPW, with a central conductor of width $w$, a separation gap to the ground planes $g$, the capacitance and inductance per unit length in the limit of zero thickness are \cite{simons2004coplanar}
\begin{align}
\mathcal{C}&=\frac{1+ \epsilon_{r}}{2}\epsilon_0 \frac{4K(k_0)}{K(k'_0)},\\
\mathcal{L}&=\frac{\mu_0}{4}\frac{K(k'_0)}{K(k_0)},
\end{align}
where $\epsilon_0$ and $\mu_0$ are the vacuum permittivity and permeability, respectively, $\epsilon_{r}$ is the relative dielectric constant of the substrate, $K$ are complete elliptic integral of the first type, with 
\begin{align}
k_0=\frac{w}{w+2g}, \hspace{1cm} k_{0}^{\prime}=\sqrt{1-k_0^{2}},
\end{align}
thus, the line impedance is 
\begin{equation}
    Z_0=\frac{377}{16}\frac{1}{\sqrt{\epsilon_{\text{eff},0}}}\frac{K(k'_0)}{K(k_0)} [\Omega],
\end{equation}
with $\epsilon_{\text{eff},0}= (1+ \epsilon_{r})/2$. The propagation velocity of the zero-thickness CPW is 
\begin{align}
    v_0=\frac{1}{\sqrt{\mathcal{LC}}}=\frac{c}{\sqrt{ \epsilon_{\text{eff},0}}},
\end{align}
where $c$ is the speed of light in vacuum. For finite metallic film thickness $t$, there are corrections to these expressions \cite{gao2008thesis}. We introduce
\begin{align*}
    d(t)&= \frac{2t}{\pi},\\
    u_1(t)&=\frac{w}{2}+\frac{d(t)}{2}+\frac{3\log2}{2}d(t)-\frac{d(t)}{2}\log\frac{2d(t)}{w} \\
    &+\frac{d(t)}{2}\log\frac{g}{w+g},\\
u_2(t)&=\frac{w}{2}+g-\frac{d(t)}{2}-\frac{3\log2}{2}d(t)+\frac{d(t)}{2}\log\frac{2d(t)}{w+2g}\\
&+\frac{d(t)}{2}\log\frac{g}{w+g},
\end{align*}
such that
\begin{align}\label{eq:Cg}
\mathcal{C}_g&=\epsilon_0 \frac{2K(k_t)}{K(k'_t)}+\epsilon_{r}\epsilon_0 \frac{2K(k_0)}{K(k'_0)},\\
\label{eq:Lg}
\mathcal{L}_g&=\frac{\mu_0}{4}\frac{K(k'_{t/2})}{K(k_{t/2})},
\end{align}
where $k_t=u_1(t)/u_2(t)$ and $k'_t=\sqrt{1-k_t^2}$. We choose geometrical values which we can fabricate accurately with optical lithography: central conductor width $w=20$ $\mu$m and gap $g=11$ $\mu$m which give $50.1\Omega$ for a $t=70$\ nm film in absence of kinetic inductance. 

When kinetic inductance is taken into account, the inductance per unit length of the CPW is modified \cite{watanabe1994CPWwithki,gao2008thesis}
\begin{align}
\mathcal{L}_t&=\mathcal{L}_g+\mathcal{L}_k,
\end{align}
with 
\begin{align}\label{EqG}
\mathcal{L}_k= \frac{L_k^{\square}}{w}G(w,g,t)
\end{align}
 depending on the penetration depth $\lambda$ ($L_k^{\square}=\mu_0 \lambda^2/t$) and the geometric factor
\begin{align*}
G(w,g,t)&=\frac{1}{2k_0^{'2}K^2(k_0)}\left[ -\log\frac{t}{4w}-\frac{w}{w+2g}\log\frac{t}{4(w+2g)}\right.\\&+\left.\frac{2(w+g)}{w+2g}\log\frac{g}{w+g}\right].
\end{align*}
Here we have used the typo correction to the expression of Watanabe \textit{et al.} pointed out by J. Clem \cite{Clem2013CPWkin}. With these expressions, it is possible to get a first estimate on the frequency of the resonators with and without kinetic inductance.

The most reliable information for the design is obtained from finite-element simulations with Qiskit Metal\cite{qmetal} and Ansys to verify our designs. As shown in Fig. \ref{fig:fig6}(d), we translate the design of panel (a) into a quasi-3D model where we give the properties of the dielectric and the metallic layer. We compute the complex transmission coefficient as a function of frequency. The results for the amplitude and phase close to a resonance are shown in (e1-3) and (f1-3), respectively. We fit these features to extract the resonance frequency and coupling quality factors. We simulate several resonators changing the total length $\ell$ and the coupling length $\ell_c$ (see panel (a)). The resulting resonance frequency as a function of $\ell$ is shown in (g). As expected, it follows a $1/\ell$ dependence, and from the inset we extract the effective dielectric constant. Since we do not include losses, the quality factor is determined exclusively by the coupling quality factor. We plot in (h) the resulting value of $Q_c$ as a function of $\ell_c$ in a situation where we adjust the total $\ell$ such that the resonance frequency remains fixed. This approach allows us to design all three resonators on each chip with the same target $Q_c$ by using the figure of merit shown in the inset. 

\section{Measurement setup}
\label{ap:meas_setup}   
\begin{figure*}[t!]
    \centering
    \includegraphics[width=\linewidth]{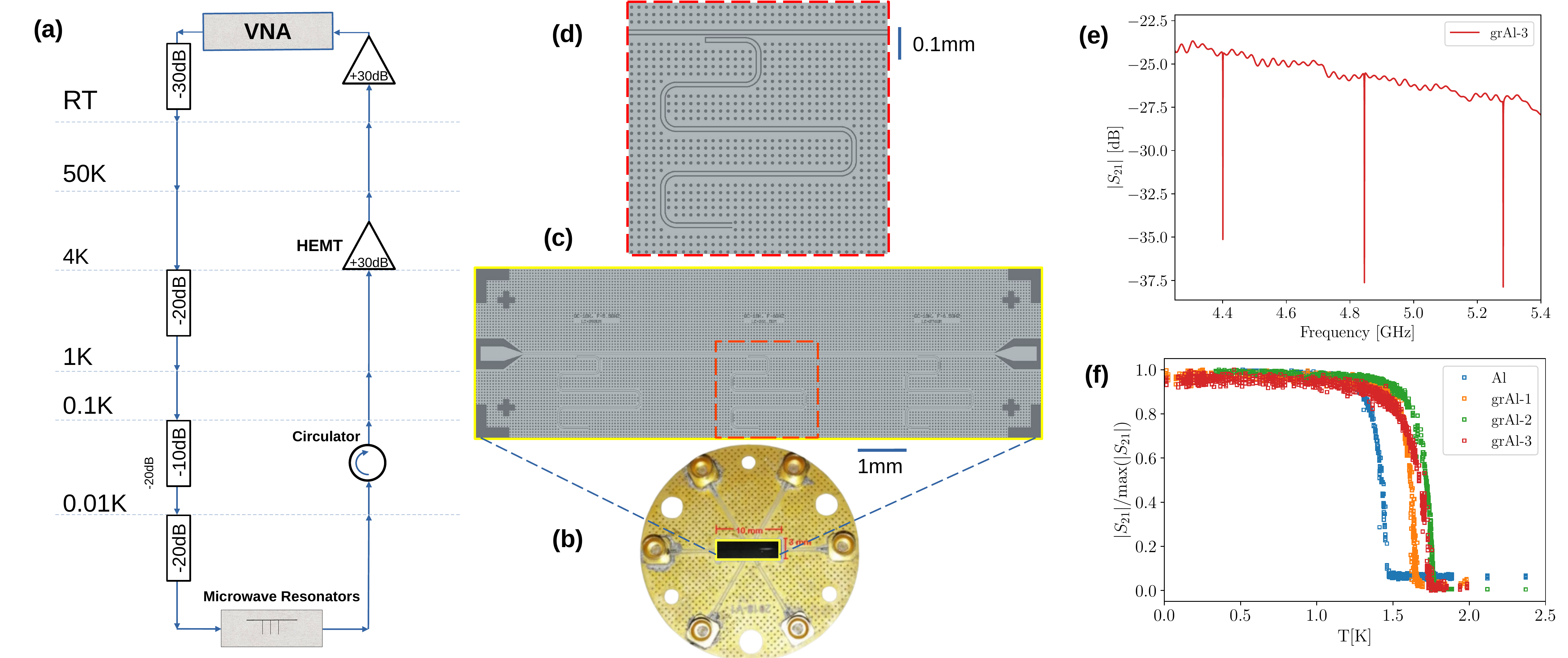}
    \caption{\textbf{Measurement Setup:} (a) Microwave setup connecting the VNA at room temperature with the sample anchored at the lowest temperature stage of a dilution refrigerator. (b) Circular sample holder equipped with Mini-SMP connectors. (c) Three microwave resonators with different frequency. (d) Quarter wavelength resonator capacitively coupled to a feedline. (e) Measured amplitude of $S_{21}$ for different temperatures to extract the $T_c$. (f) Transmission measurement versus frequency for the grAl-3 sample at 10\ mK.}
    \label{fig:setup}
\end{figure*}
We use a vector network analyzer (VNA) to measure the transmission coefficient and probe the resonators (see Fig \ref{fig:setup} (a)). The output signal is first attenuated by -30dB at room temperature and as the signal travels through the cryostat, it undergoes an additional total attenuation of -50dB with attenuators thermalized at different temperatures. After probing the microwave resonators, the signal passes through a circulator/isolator and is subsequently amplified by a high electron mobility transistor (HEMT) amplifier located at 4K. The circulator serves to isolate the sample from the noise of the amplifier. Finally, the signal is further amplified at room temperature before being directed to the input of the VNA. 

The samples are diced to dimensions of 10mm x 3mm (see Fig \ref{fig:setup} (b)) and then mounted and wire-bonded onto a circular sample holder equipped with Mini-SMP connectors, which are subsequently connected to the SMA connectors at the mixing chamber stage of the cryostat. Each sample contains an array of three microwave resonators (Fig \ref{fig:setup} (c)), designed to operate at different resonance frequencies. All resonators are capacitively coupled to a single transmission line, as illustrated in Fig  \ref{fig:setup}(d). The resulting $|S_{21}|(f)$ in a large frequency scale is shown in (e). 

\section{Complementary microwave characterization of grAl resonators}
\label{ap:extra}
The initial characterization of grAl films involves measuring the superconducting transition temperature $T_c$. At a fixed microwave frequency, we measure the amplitude of the transmission coefficient at various temperatures of the films. Fig \ref{fig:setup}(f) shows the normalized transmission coefficient at different temperatures. We observe $T_c$ increases with the resistance of each film. 

\section*{Acknowledgments}
We are thankful to Louise Noell for her engagement at the early stages of this project during her internship (Instituto Balseiro - École PHELMA, Grenoble). We thank Hélène Le Sueur and Pief Orfila from the Quantronics Group (France) for the training in the growth of granular aluminium. We thank the team of the Clean Room in Centro Atómico Bariloche for the help in the fabrication. We are also grateful to Juan Díaz for his assistance with the DC resistance measurements of our samples. We thank Claudio Ferrari, Andrés Di Donato and Juan Bonaparte from the DMNT at Centro Atómico Constituyentes for fabricating our optical lithography masks. We thank Luis Avilés, Andres Hofer, Javier Gómez and Néstor  Haberkorn for their support with the sputtering facility. We are thankful to Martin Sirena for the AFM measurements, to Bernardo Pentke for the SEM pictures of our films, to Carlos Bertoli for the FIB service and to Adriana Condo for the TEM images. We thank Pablo Pedrazzini for helping us with DC transport measurements when our fridge was not yet installed. We thank Vjeko Dimi\'c, Alex Kirchner, Ayelén Prado, Diego Pérez and Mariano Real for carefully reading the manuscript and for the useful discussions. We acknowledge support from Instituto Balseiro, CNEA, CONICET, PIP 11220200101825CO TOSI. L.T. acknowledges the Georg Forster Fellowship from the Alexander von Humboldt Stiftung, Germany. 

\bibliography{biblio.bib}

\end{document}